\documentclass[12pt,preprint]{aastex}

\newcommand{\etal}{{\em et al.}}            

\shorttitle{Hot Gas Environment of 3C\,388}
\shortauthors{Kraft \etal}

\begin{document}

\title{The Hot Gas Environment of the Radio Galaxy 3C\,388: Quenching the Accumulation of Cool Gas in a Cluster
Core by a Nuclear Outburst}
\author{R. P. Kraft}
\affil{Harvard/Smithsonian Center for Astrophysics, 60 Garden St., MS-67, Cambridge, MA 02138}
\author{J. Azcona}
\affil{Observatoire de Paris-Meudon-Nancy, 5 Place Jules Janssen, 92195 Meudon Cedex, France}
\author{W. R. Forman}
\affil{Harvard/Smithsonian Center for Astrophysics, 60 Garden St., MS-2, Cambridge, MA 02138}
\author{M. J. Hardcastle}
\affil{University of Hertfordshire, School of Physics, Astronomy, and Mathematics, Hatfield AL10 9AB, UK}
\author{C. Jones, S. S. Murray}
\affil{Harvard/Smithsonian Center for Astrophysics, 60 Garden St., MS-2, Cambridge, MA 02138}

\begin{abstract}

We present results from a 35 ks {\em Chandra}/ACIS-I observation of the
hot ICM around the FR II radio galaxy 3C 388.
3C 388 resides in a cluster environment with an ICM temperature of $\sim$3.5 keV.
We detect cavities in the ICM coincident with the radio lobes.
The enthalpy of these cavities is $\sim$1.2$\times$10$^{60}$ ergs.
The work done on the gas by the inflation of the lobes is $\sim$3$\times$10$^{59}$ ergs,
or $\sim$0.87 keV per particle out to the radius of the lobes.
The radiative timescale for gas at the center of the cluster at the current
temperature is a few Gyrs.
The gas in the core was probably cooler and denser before the outburst,
so the cooling time was considerably shorter.  We are therefore
likely to be witnessing the quenching of a cluster cooling flow by a radio galaxy
outburst.  The mechanical power of the lobes is at least
20 times larger than the radiative losses out to the cooling radius.
Outbursts of similar power with a $\sim$5\% duty cycle would be more
than sufficient to continually reheat the cluster core over the Hubble
time and prevent the cooling of any significant amount of gas.
The mechanical power of the outburst is also roughly two orders of magnitude
larger than either the X-ray luminosity of the active nucleus
or the radio luminosity of the lobes.
The equipartition pressure of the radio lobes is more than an order
of magnitude lower than that of the ambient medium, indicating that
the pressure of the lobe is dominated by something other than the
relativistic electrons radiating at GHz frequencies.

\end{abstract}


\section{Introduction}

Our understanding of radio plasma/ISM interactions in radio
galaxies has been revolutionized by
{\em Chandra} X-ray Observatory studies of the gaseous atmospheres surrounding
these sources.  {\em Chandra} is the first X-ray observatory with
spatial resolution sufficient to carry out detailed studies of the shells,
bow shocks, bubbles, cavities, cold filaments, and surface
brightness discontinuities in the gas that are the consequences
of powerful AGN outburst.
X-ray studies of the gaseous environments of radio galaxies
give us important clues about the dynamics, the energetics, and
the temporal evolution of AGN outbursts
that cannot be directly obtained
from radio observations.

Interest in radio galaxy/ICM interactions has flourished in the {\em Chandra} era
after the realization that the large cluster cooling flows predicted
by the earlier generation of X-ray observatories do not exist \citep{pet01,kaa04}.
Since the radiative cooling time of the gas at the center of many clusters
is a few Gyrs or less, the gas must be reheated, either occasionally or
continuously, to prevent the formation of large cooling flows.  There are several possible
sources of energy to reheat the core including cosmic rays \citep{boh88},
thermal conduction from the hot gas halo \citep{tuc83, dol04} 
or supernovae \citep{mcn04}.
The most promising method to reheat cooling cluster cores and
balance radiative losses is heating by nuclear outbursts \citep{tab93, chu02}.
In this scenario, there is a cyclical relationship between the cooling gas
and the formation of radio galaxies.  As the gas cools, it falls in toward the
central supermassive black hole, initiating a nuclear outburst.  The mechanical
energy of the nuclear outflow heats (either via shocks or via conversion of bubble
enthalpy to thermal energy of the ambient gas) the ICM \citep{rey01,rey02}.
AGN outbursts therefore play a critical role in the evolution of
clusters of galaxies.  Occasional outbursts can provide
sufficient energy to reheat the radiatively cooling cores of galaxy
clusters and prevent large cooling flows from forming.

The radio galaxy 3C 388 is classified as a Fanaroff-Riley type
II (FR II) radio galaxy, although its 
luminosity ($P_{178\ MHz}$=5$\times$10$^{25}$ W Hz$^{-1}$ \citep{fan74})
lies near the FR I/II dividing line.
The radio morphology of this source
is closer to a `fat-double' \citep{owe89}
than the canonical FR II `classical double' such as Cyg A and 3C 98.
Optically, the nucleus of 3C 388 is classified as a low-excitation
radio galaxy \citep{jac97}.
Multi-frequency VLA observations of 3C 388 show significant structure
in spectral index maps which has been interpreted as evidence
for multiple nuclear outbursts \citep{roe94}.
Previous X-ray observations of this radio galaxy have shown that
it is embedded in a cluster environment \citep{fei83, har99, lea01}.
The local galaxy environment is extremely dense \citep{pre88}, and the
central elliptical galaxy which hosts 3C 388
is one of the most luminous ($M_B$=-24.24) in
the local Universe \citep{owe89, mar99}.

In this paper we present results from a 35 ks observation of the hot
gas around 3C 388.  This paper is organized as follows.  Section two
contains a summary of the observational details.  
The results of the data analysis are presented in section 3.
We discuss the implications of our results in section 4.
Section 5 contains a brief summary and conclusions, as well
as possible future observations.
We assume WMAP cosmology throughout this paper \citep{spe03}.  The observed redshift
($z$=0.0908) of the host galaxy of 3C 388
corresponds to a luminosity distance of 410.3 Mpc, and one arcminute
is 100.32 kpc.  All uncertainties are at 90\%
confidence unless otherwise stated, and all coordinates are J2000.
All elemental abundances in this paper are relative to the Solar
value tabulated by \citet{angr}.
Absorption by gas in our Galaxy ($N_H$=6.32$\times$10$^{20}$ cm$^{-2}$) is included in
all spectral fits \citep{dic90}.

\section{Observations}

\begin{figure}
\plotone{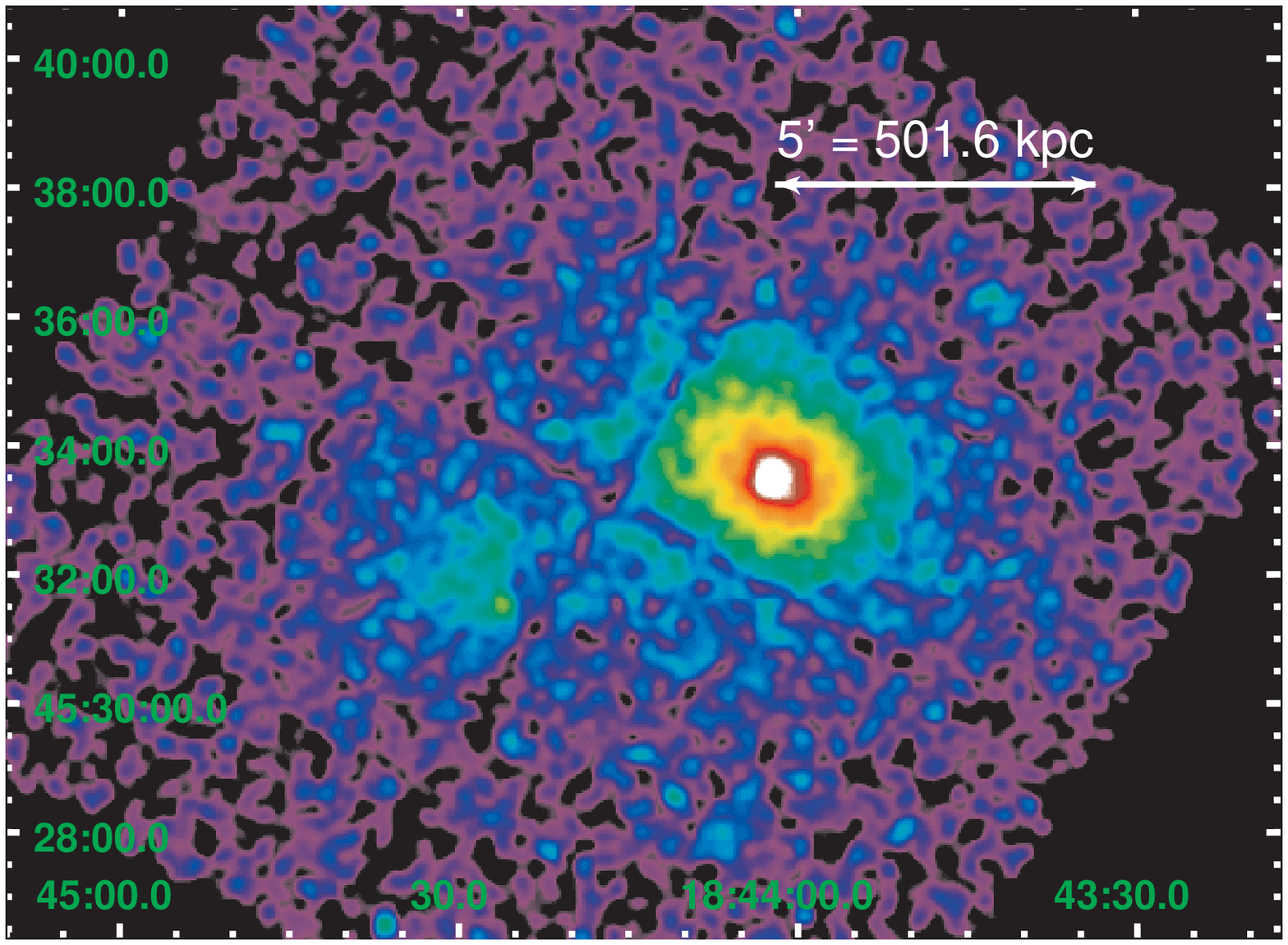}
\caption{Gaussian smoothed (6$''$ r.m.s.) {\em Chandra}/ACIS-I
image of 3C 388 in the 0.5-2.0 keV band with point sources removed.}\label{gsmooth}
\end{figure}

The radio galaxy 3C 388 was observed twice with {\em Chandra}/ACIS-I in VF mode.  The
first observation was made on January 9, 2004 (OBSID 4765) for 7.5 ks,
and was terminated because of high background.  The second
observation was made on January 29, 2004 (OBSID 5295) for 32.5 ks.
In this paper we present data only from the second observation.
The complications of combining the data sets (different instrument roll angles and
higher background in the first observation) more than offset the
advantage of the additional $\sim$25\% observing time.
We made light curves for each CCD in the 5.0 to 10.0 keV band to search
for background flares and removed intervals where the background rate
was more than 3$\sigma$ above the mean,  leaving 30717.5 s of good data.
VF filtering was applied to the data to reduce the background.
Bad pixels, hot columns, and columns along node boundaries also were removed,
and standard ASCA grade filtering (0,2,3,4,6) was applied to the events file.
Diffuse emission from the ICM fills the {\em Chandra} FOV in this
observation, so it was not, in general, possible to use a local background measured
in the same observation.  Background for both imaging and spectral
analysis was estimated using the high latitude deep sky observations
provided by the CXC.
An image of the ACIS-I FOV smoothed by a Gaussian (6$''$ r.m.s.) in the 0.5-2.0
keV band is shown in Figure~\ref{gsmooth}.

We use archival VLA observations of 3C 388 at two frequencies, 8.4 GHz and 1.4 GHz, to
make X-ray/radio comparisons.
Radio spectral index maps of \citet{roe94} show two spatially separated
regions in each lobe.  Near the radio peaks of each lobe, the spectral
indices of the radio flux density ($S_\nu\propto\nu^{-\alpha}$) are
between 0.6-0.8.  The radio spectral index of the more distant regions
is considerably steeper ($\sim$1.5), with a sharp discontinuity defining
these distinct regions.  \citet{roe94} argued that the radio lobe is thought to be
inflating into the relic lobe of a previous outburst.

\section{Analysis}

\begin{figure}
\plotone{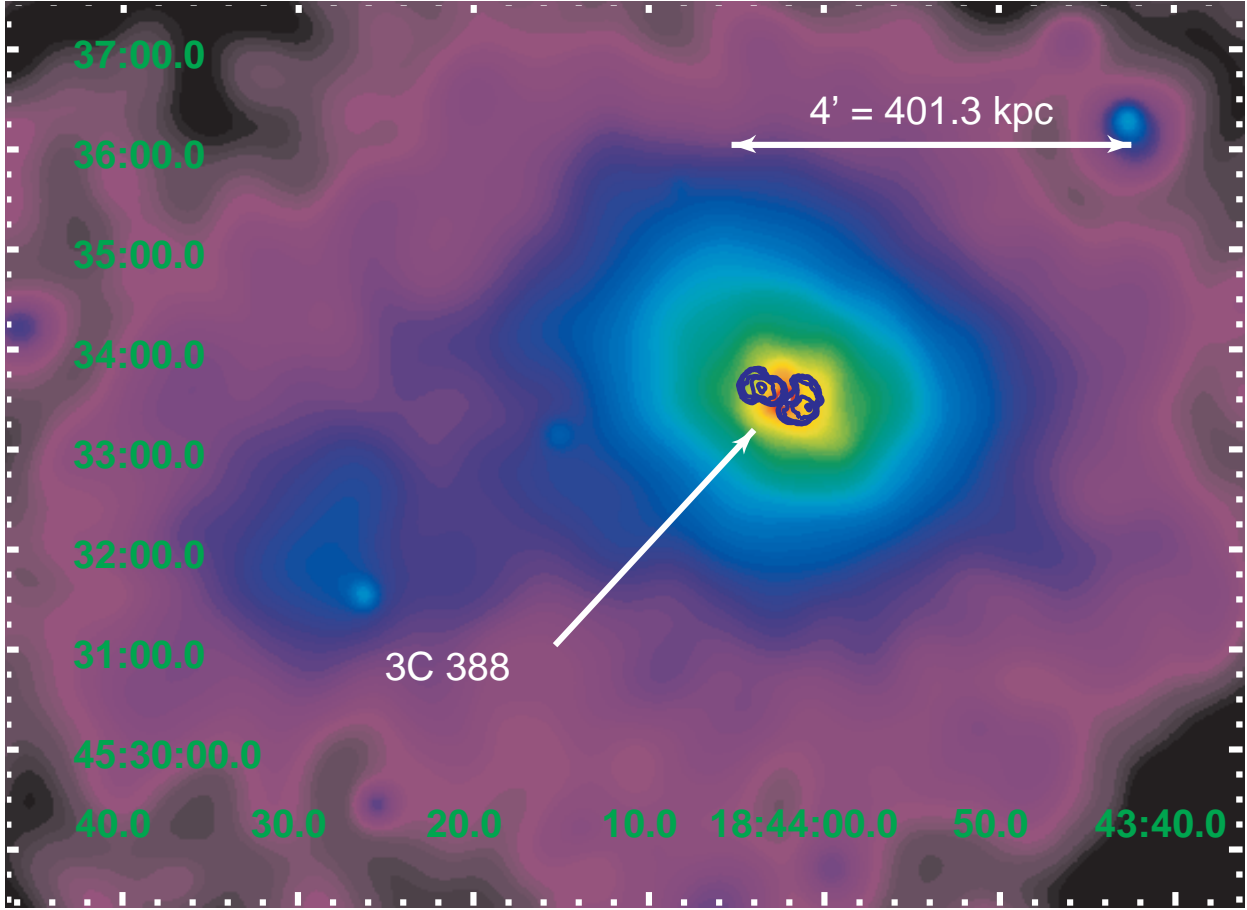}
\caption{Adaptively smooth, background subtracted, exposure corrected {\em Chandra}/ACIS-I
image of 3C 388 in the 0.5-2.0 keV band with 1.4 GHz radio contours
overlaid (blue).  The bright nucleus, ICM, and southeastern
concentration are clearly visible.}\label{cluster}
\end{figure}

An adaptively smoothed, exposure corrected, background subtracted
{\em Chandra}/ACIS-I image of 3C 388 in the 0.5-2.0 keV band is shown
in Figure~\ref{cluster}.  All point sources with four or more counts, other than the
active nucleus of 3C 388, have been removed and replaced by the mean background
from an adjacent annulus.
Diffuse emission from the cluster gas fills the ACIS-I FOV.
Contours from the 1.4 GHz radio map of 3C 388 are overlaid.
The radio source is small (1$'$/100 kpc across) and lies at the center of the cluster.  
A secondary peak of X-ray emission lies
$\sim$4.8$'$ (482 kpc) east of the cluster center, and is perhaps associated
with a merging subcluster.
The X-ray isophotes within $\sim$1$'$ of the nucleus are
elliptical ($e\sim$0.3).  The major axis of the elliptical isophotes
is aligned along the axis of the radio jet, suggesting that the NE/SW
extension may be related to the nuclear outburst.  On larger scales, however, the isophotes
are approximately circular (ignoring the peak to the East).
It is likely that the observed ellipticity of the X-ray isophotes is
the result of the inflation of the radio lobe
as observed in simulations \citep{bas03}.  For simplicity we assume spherical symmetry
in our analysis below.

\subsection{Radio Lobe/ICM Interaction}

\begin{figure}
\plotone{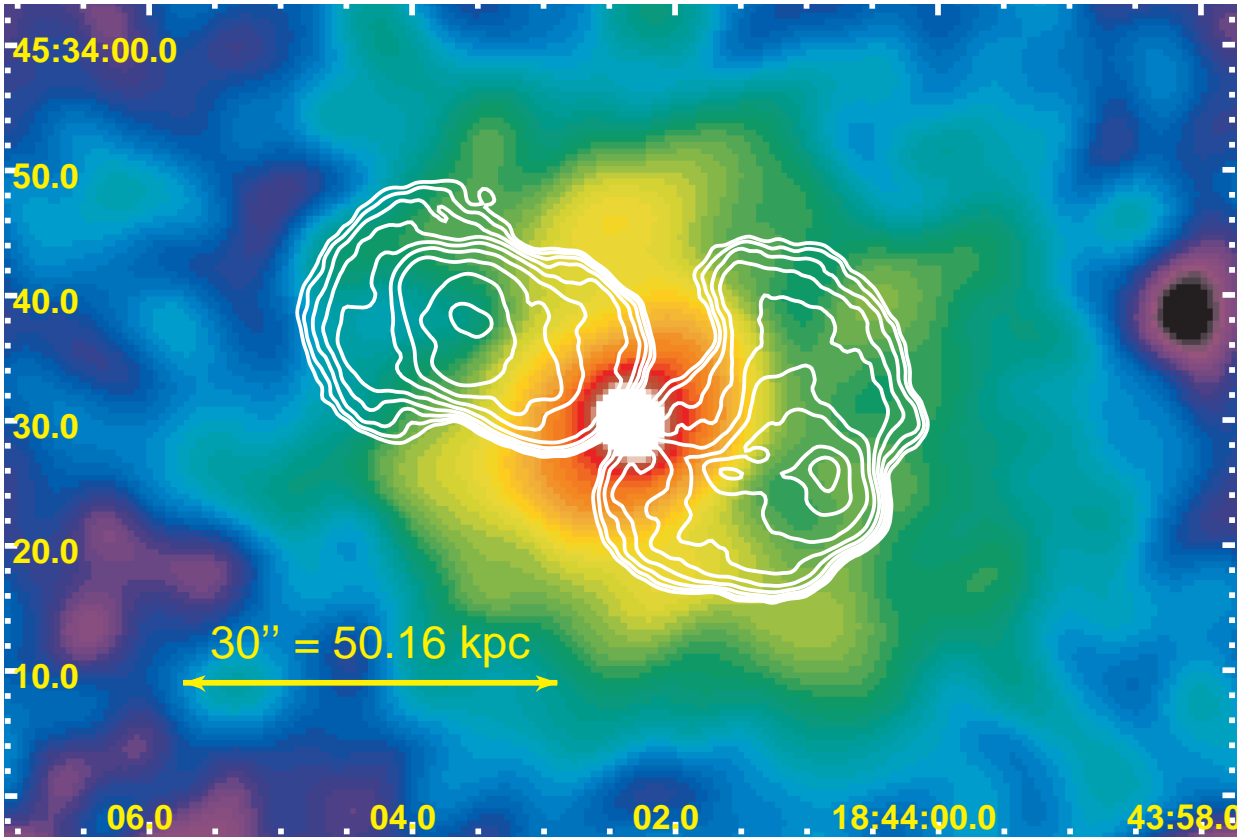}
\caption{Adaptively smooth, background subtracted, exposure corrected {\em Chandra}/ACIS-I
image of 3C 388 in the 0.5-2.0 keV band with 5 GHz radio contours overlaid.}\label{center}
\end{figure}

\begin{figure}
\plotone{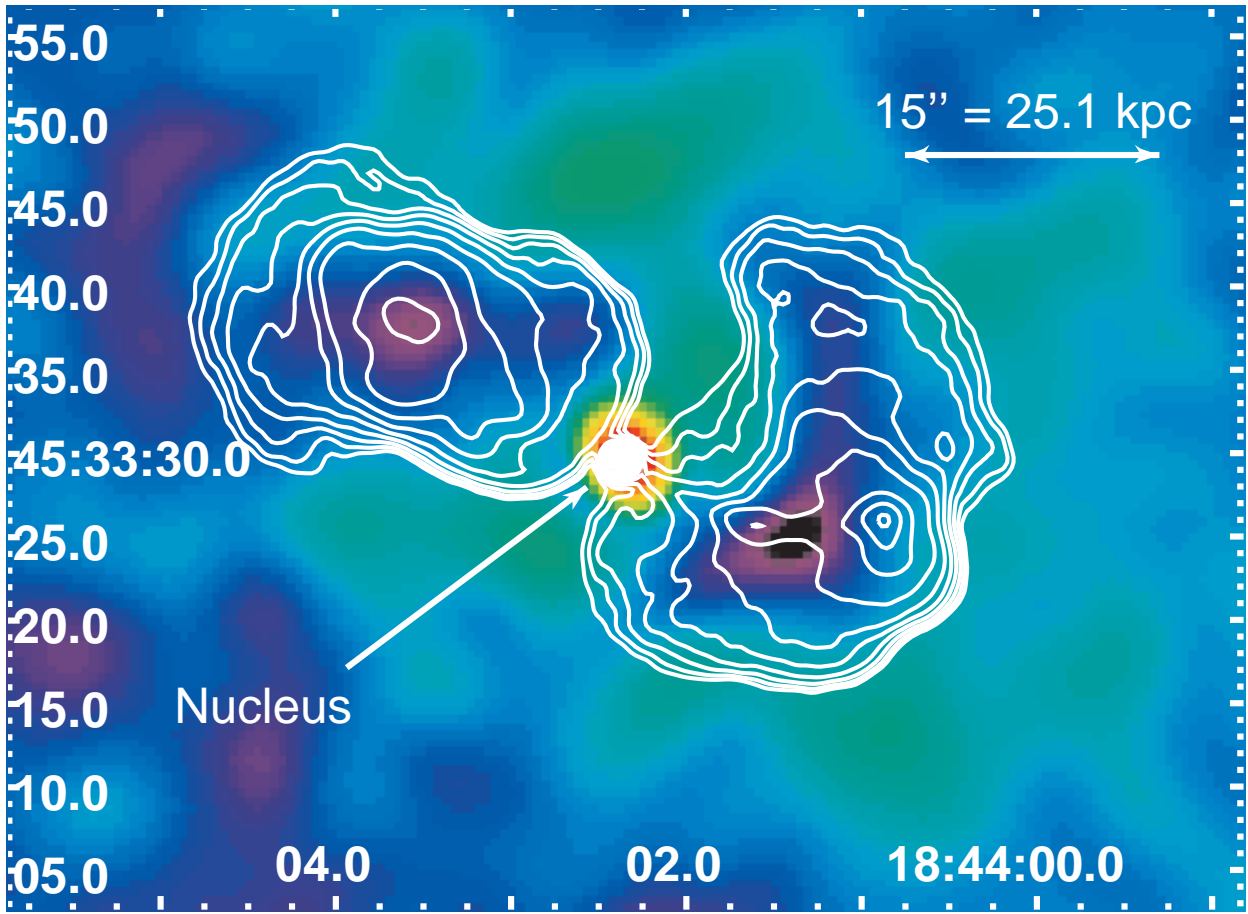}
\caption{Smoothed (2$''$ Gaussian r.m.s.), background subtracted, exposure corrected {\em Chandra}/ACIS-I
image of 3C 388 in the 0.5-2.0 keV band with azimuthal profile subtracted
to enhance the appearance of the
X-ray cavities associated with the radio lobes.  Radio contours (1.4 GHz) are
overlaid.  The `peaks' of the surface brightness decrements (shown as black at the center
of the lobes) correspond to a $\sim$25\% decrease in the X-ray
surface brightness relative to the azimuthal average.}\label{resid}
\end{figure}

Figure~\ref{center} contains an adaptively smoothed, background
subtracted, exposure corrected {\em Chandra}/ACIS-I image of the
central 1.5$'$$\times$1.1$'$ region with 8.4 GHz radio contours
overlaid.  The X-shaped X-ray morphology, typical of
radio plasma/ICM interactions (e.g. M84 \citep{fin01} and NGC 4636 \citep{jon02}),
is clearly visible in Figure~\ref{center}.
There are decrements in the X-ray surface brightness coincident
with the radio lobes.
To enhance the visibility of these decrements, we have subtracted 
the azimuthally averaged surface brightness
profile from Figure~\ref{center}.  The residual image is shown in Figure~\ref{resid}.
The inflation of the radio lobes has evacuated cavities
in the gas and created the deficits in the X-ray surface brightness.

\begin{figure}
\plotone{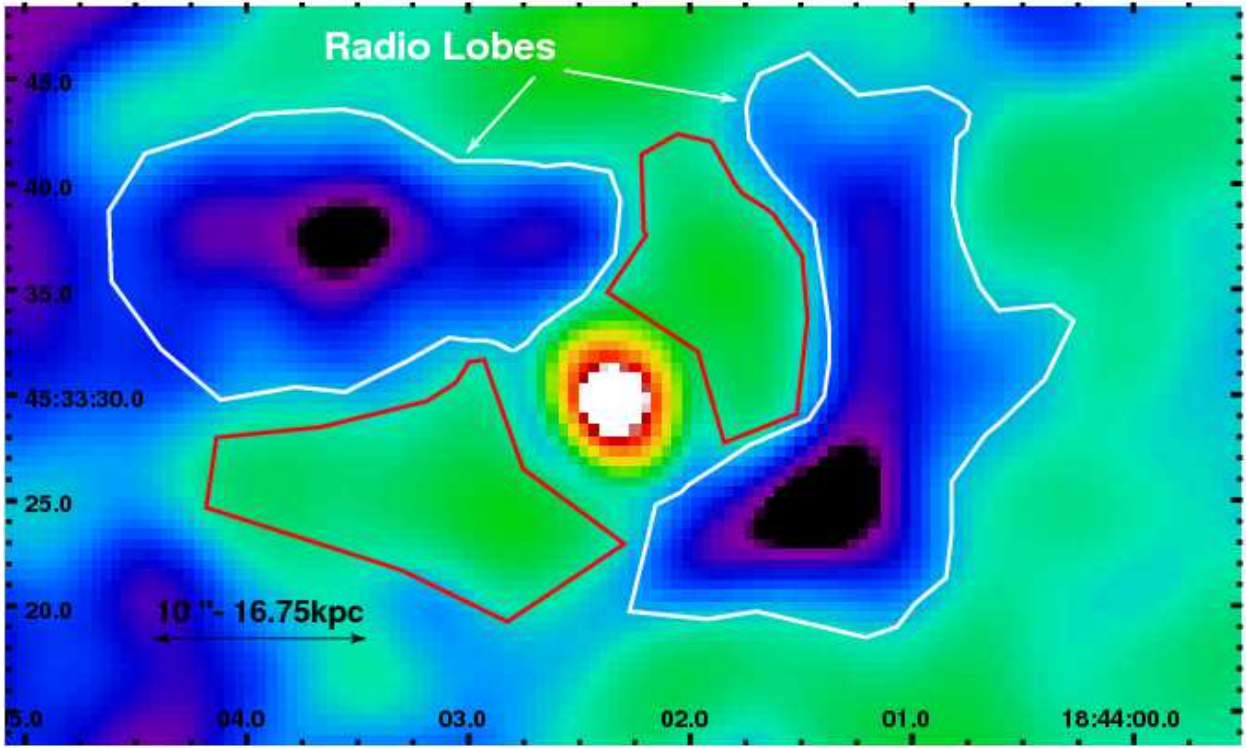}
\caption{Regions extracted for spectral analysis.  Region 1 is
coincident with the radio lobes (white), region 2 is interior
to the lobes (red), and region 3 is an elliptical annulus
with inner and outer semi-major axes of 25$''$ and 50$''$, respectively.
The best fit temperatures and abundances with 90\% uncertainties
are summarized in Table~\ref{spectab}.}\label{central}
\end{figure}

To constrain the dynamics of the interaction of the radio lobes
with the ICM, we fit the X-ray spectra in three groups of regions
in the vicinity of the lobes.  Two of the groups are shown in Figure~\ref{central}.
The regions outlined in red (combined for spectral analysis) represent
the gas in the central region of the cluster, and the two regions in
white are coincident with the X-ray cavities created by the inflation
of the lobe.  A third region (not shown on Figure~\ref{central}) is an
elliptical annulus with inner and outer major axes of 25$''$ and 50$''$,
respectively, with the major axis aligned along the jet.
The spectra were fit using a single temperature APEC model and
Galactic absorption.  The best fit temperatures and elemental abundances
for each region are summarized in Table~\ref{spectab} (uncertainties are
at 90\% confidence).  There is no statistically significant evidence
for any temperature structure in the regions chosen, although
the uncertainties are large enough that the presence of a weak
shock cannot be excluded.
The absence of gas hotter than the ambient ICM demonstrates
that the lobes are not expanding with large Mach number.  In addition,
there is no evidence for cool rims similar to those seen around cavities in
other clusters (e.g. Hydra A \citep{mcn00}) that are thought
to have been dredged up from the center of the cluster.

\subsection{Large Scale X-ray Emission}

The azimuthally averaged radial surface brightness profile of the
X-ray emission from the gas is shown in Figure~\ref{sbprof}.  The best-fit
isothermal $\beta$-model profile has been overlaid.
We find $\beta$=0.444$\pm$0.003 and a core radius $r_0$=8.30$''$$\pm$0.01$''$ from
fitting the surface brightness profile between
5$''$ and 200$''$ from the nucleus, roughly consistent with values
previously reported by \citet{lea01}.
There is an excess of emission in the central 5$''$ over the best-fit
$\beta$-model.  This is due to a combination of cooling
gas in the galaxy core and a contribution from the central AGN (see below).

\begin{figure}
\plottwo{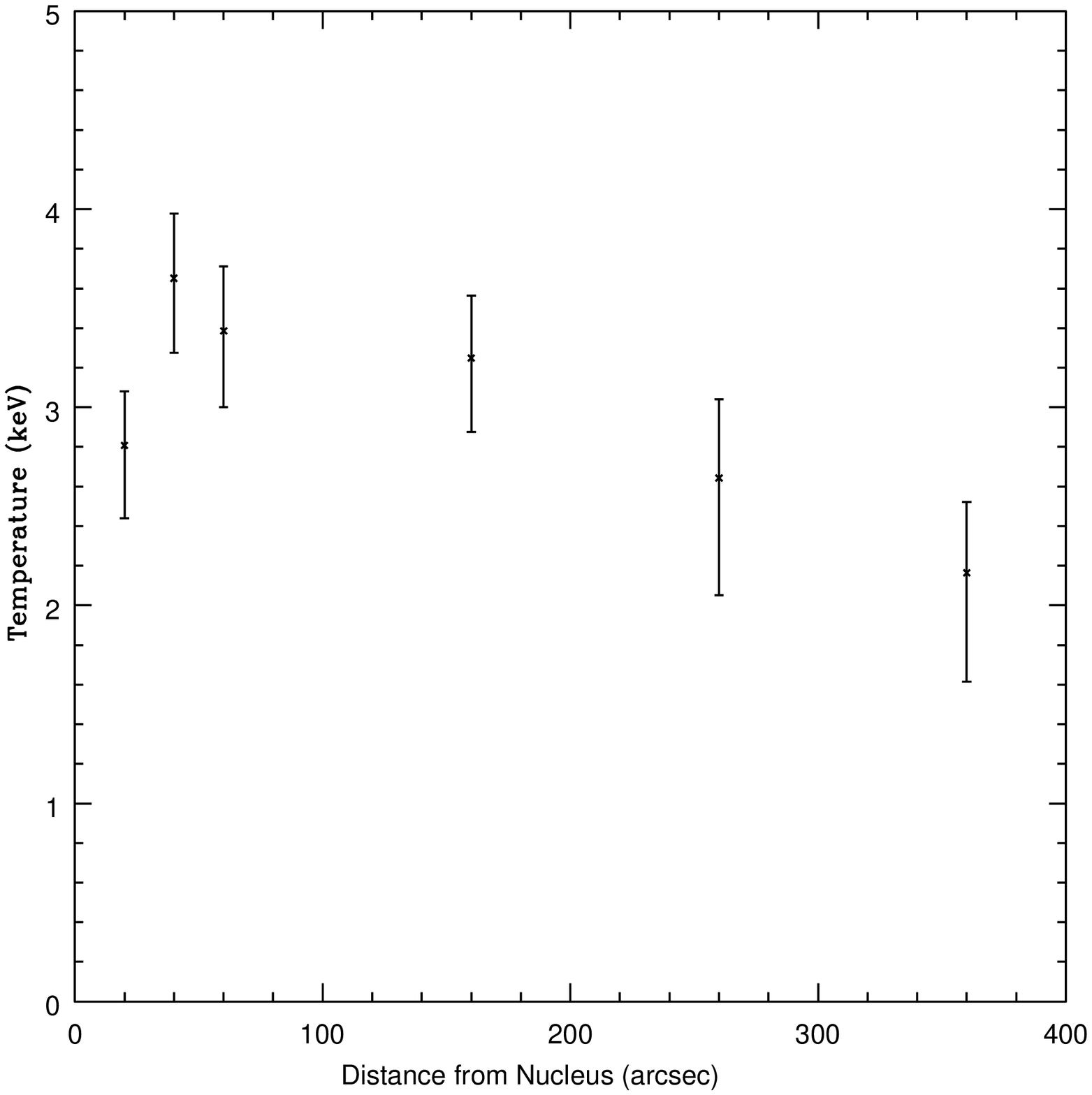}{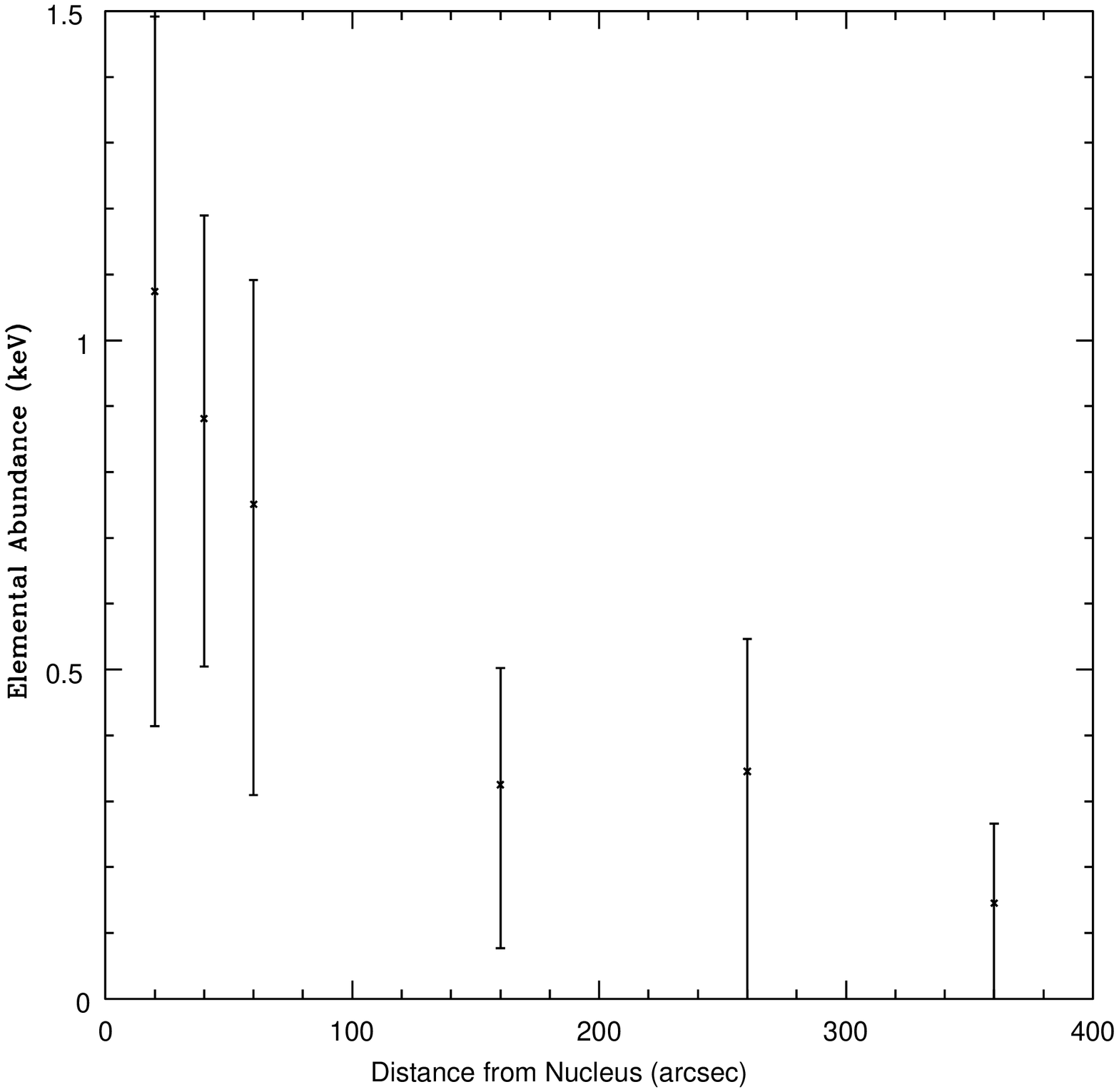}
\caption{Radial temperature and elemental abundance 
profiles of the ICM around 3C 388 in azimuthal regions.}\label{tprof}
\end{figure}

\begin{figure}
\plotone{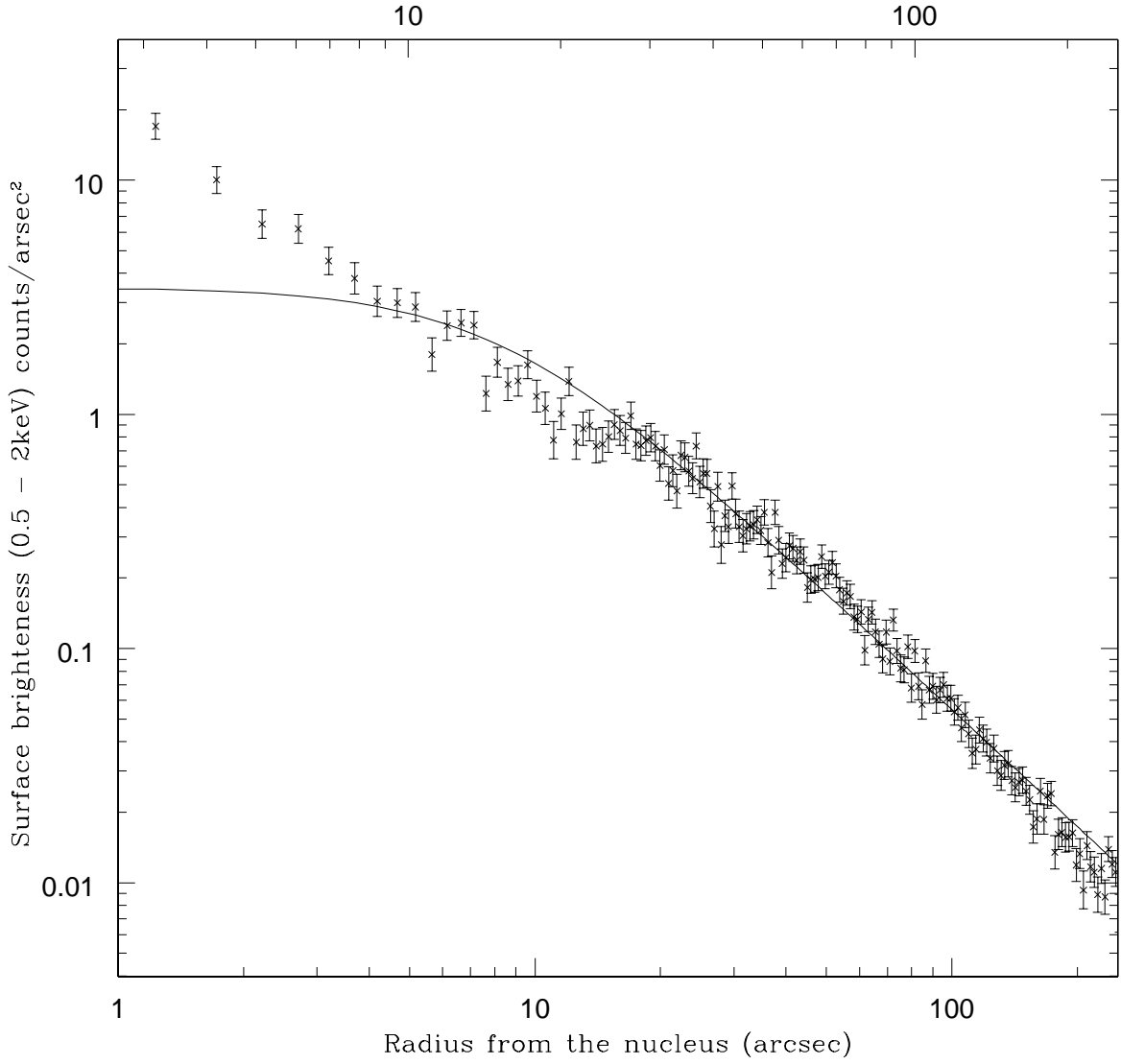}
\caption{Radial surface brightness profile of ICM around 3C 388.}\label{sbprof}
\end{figure}

\begin{figure}
\plotone{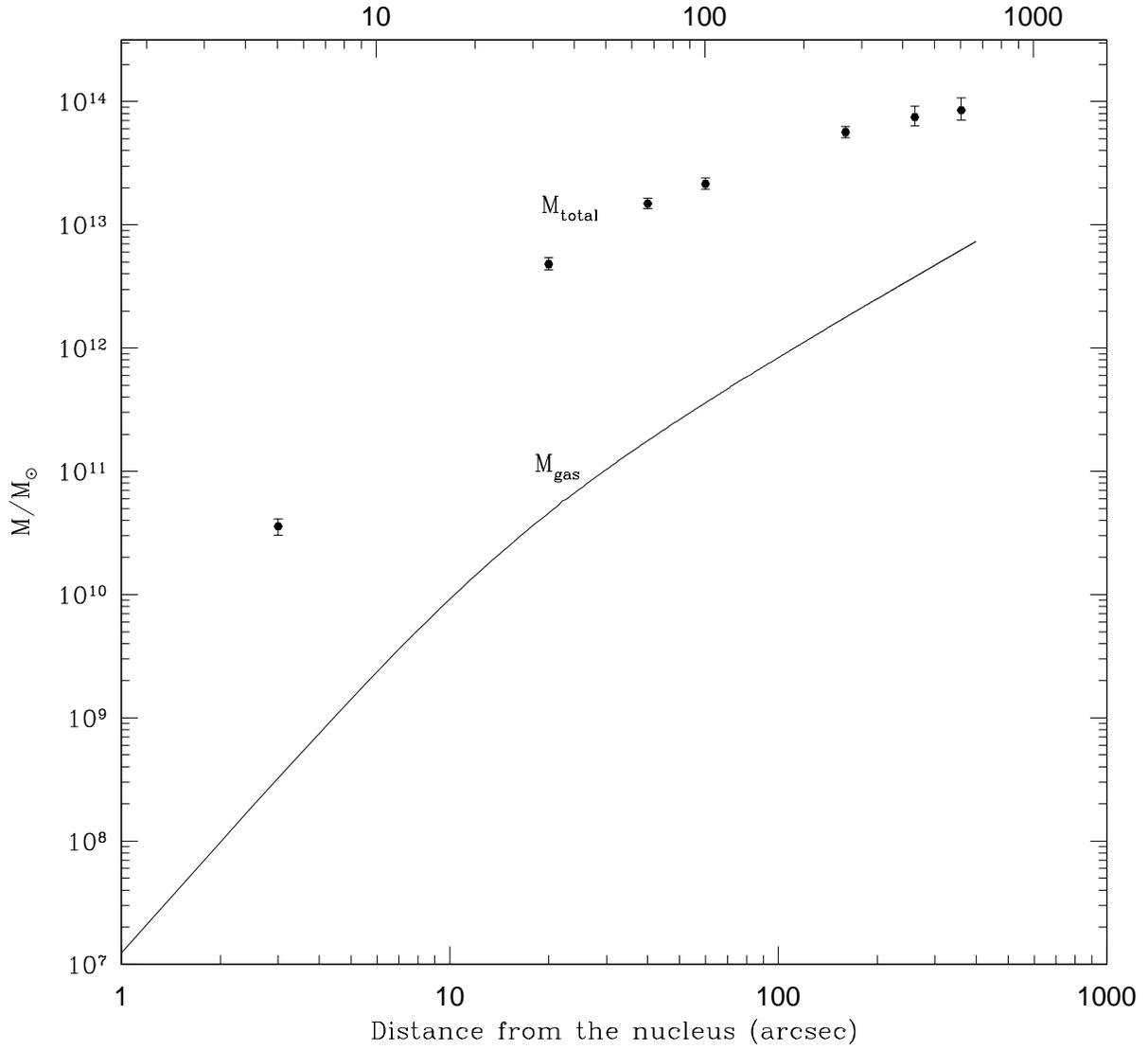}
\caption{Gas mass and gravitating mass as a function of distance from
the center of the cluster.}\label{gravmass}
\end{figure}

We fit the temperature in six annuli centered on 3C 388 using an APEC
model with Galactic absorption ($N_H$=6.32$\times$10$^{20}$ cm$^{-2}$),
excluding the subcluster to the East.
The temperature and elemental abundance were free parameters in these fits.
Plots of the temperature and elemental abundance as a function of
radius from the center of the cluster are shown in Figure~\ref{tprof}.
The temperature of the gas within $\sim$30$''$ of the nucleus (i.e.
within the boundary of the radio lobes) is $\sim$ 3 keV.  The gas
temperature rises slightly beyond the radio lobes to 3.7$\pm$0.4 keV, with
a weak decline ($T(R)\propto R^{-0.2\pm0.05}$ at 90\% confidence) as a function of 
projected distance, $R$, from the nucleus.
The X-ray luminosity of the gas within a projected radius of 401.3 kpc (30\% of
the virial radius \citep{evr96}) is
9.0$\times$10$^{43}$ ergs s$^{-1}$ in the 0.1-10.0 keV band (unabsorbed) assuming
the gas is isothermal (3.4 keV - the best-fit emission weighted temperature in this
region).  With the assumption that the gas on large scales
is in hydrostatic equilibrium with the gravitating dark matter,
the gas and gravitating masses as a function of distance from the
center of the cluster are shown in Figure~\ref{gravmass}.

\begin{figure}
\plotone{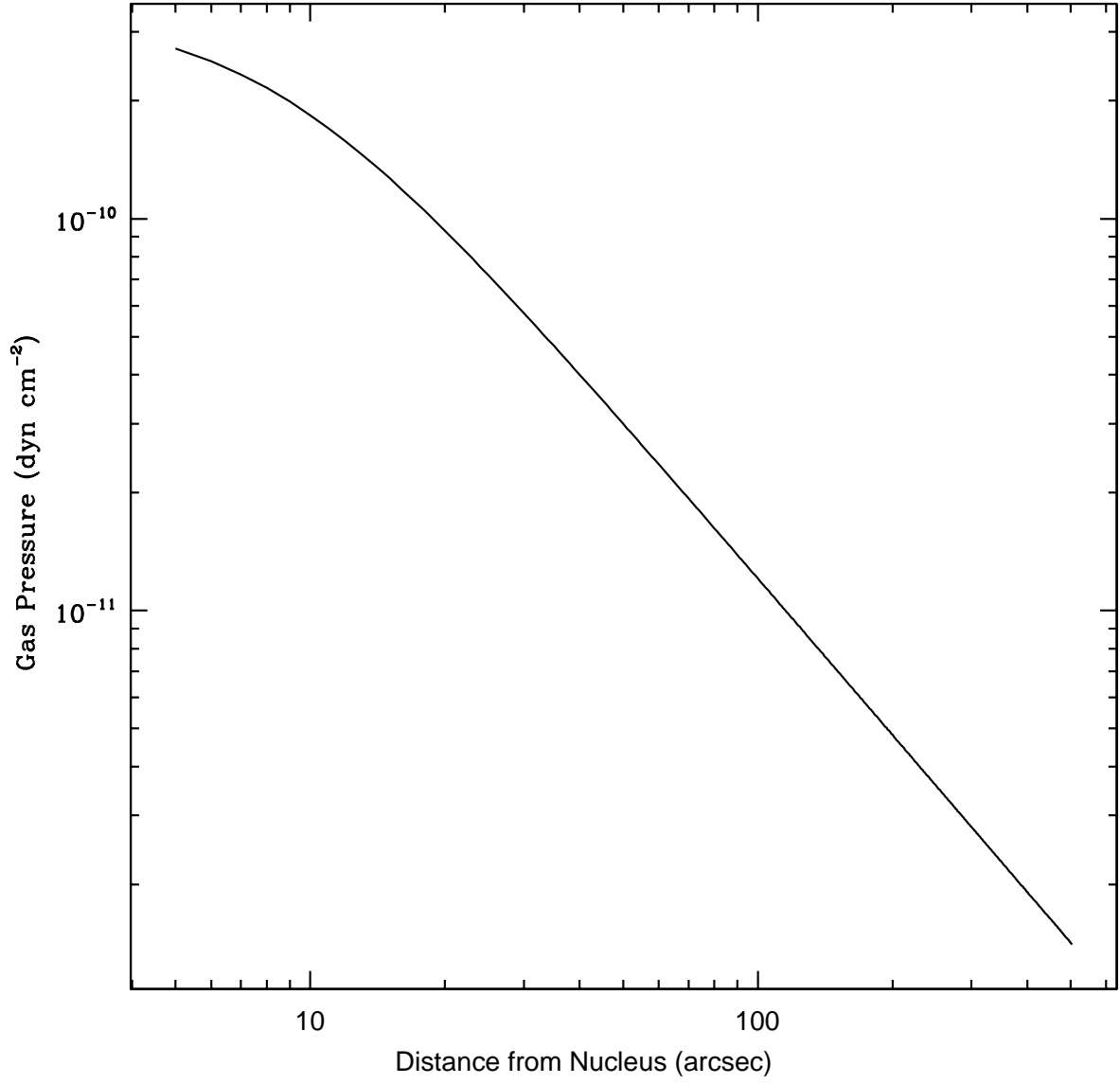}
\caption{Gas pressure of the ICM as a function of distance from the nucleus for
isothermal $\beta$-model profile.}\label{presprof}
\end{figure}

We attempted to deproject the temperature and surface brightness profiles using
the {\em projct} model of XSPEC in order to determine the temperature
and density as a function of distance from the nucleus.  However, the statistical
quality of these data was not sufficiently high to permit a detailed deprojection
in this manner.  We found that the deprojected temperature, density, and
pressure profiles between 10$''$ and 500$''$ are statistically identical to 
the isothermal ($k_BT$=3.4 keV, $Z$=0.5) $\beta$-model profile.  For simplicity,
we will use the isothermal $\beta$-model profile parameters for energy and
pressure comparisons with the radio lobes.
The central hydrogen density, $n_0$, of the isothermal $\beta$-model profile is
8.3$\times$10$^{-2}$ cm$^{-3}$.  The gas pressure as a function of distance
from the nucleus is shown in Figure~\ref{presprof}.

We also made surface brightness profiles in four wedges, 
along the two jet axes (45$^\circ$ opening angle) and the two orthogonal 
axes (90$^\circ$ opening angle), to search for a surface
brightness discontinuity that would indicate a detached shock.
Such discontinuities have been observed in {\em Chandra} observations
of the gaseous coronae of several radio galaxies including Hydra A \citep{nul05},
M87 \citep{for05}, and MS 0735.6 + 7421 \citep{mcn05}.
We find no statistically significant discontinuities in the
surface brightness distribution along any of the axes.

\subsection{Compact Components}

A point source with an unusually soft spectrum is
coincident with the active nucleus.  We fit a power law model with Galactic
absorption in the 0.5-2.0 keV band to the events within
2$''$ of the nucleus (there are few source
counts above 2 keV or below 0.5 keV).
The best-fit photon index is 2.9$\pm$0.6 (90\% confidence).  
Background was estimated in two ways: from a distant region and
in an annulus around the nucleus.
The results are insensitive to the choice of background.
The unabsorbed luminosity is 2.7$\times$10$^{42}$ ergs s$^{-1}$
in the 0.1-10.0 keV band.  
The X-ray to radio flux ratio lies along the correlation found
for samples of radio galaxy cores \citep{can99,eva05}.
We also fit an APEC model with Galactic
absorption to investigate whether the emission peak at the nucleus
could be cold gas from the galaxy with high elemental abundance.
Both the temperature and elemental abundance were free
parameters in this fit.  The thermal model was rejected at $>$99\% confidence.

No X-ray emission is detected from the jet or the compact hotspot
of the western lobe.  The 3$\sigma$ upper limit to the flux density
of the western hotspot is 5.5$\times$10$^{-16}$ ergs cm$^{-2}$ s$^{-1}$ keV$^{-1}$ (0.23 nJy)
at 1 keV, assuming a power-law spectrum with photon index 2.0.
The lower limit to the radio-to-X-ray flux density spectral index is 
$\alpha_{rx}>$ 1.06 based on an extrapolation from the 8.4 GHz flux density to
the X-ray upper limit.
The upper limit to the X-ray flux is two orders of magnitude above the
expected value for inverse Compton scattering of CMB photons or from
the SSC process assuming equipartition \citep{hard98}.
The limit on the X-ray to radio flux ratio of the hotspot is a factor
of ten below that measured for the detected 3C 403 E hotspot \citep{hard04,kra05}.
There is thus not an unusually large population of ultra-relativistic, X-ray
synchrotron emitting particles generated in this hotspot, which is consistent
with what has been observed in the majority of FR II hotspots.

\subsection{Sub-cluster merger}

\begin{figure}
\plotone{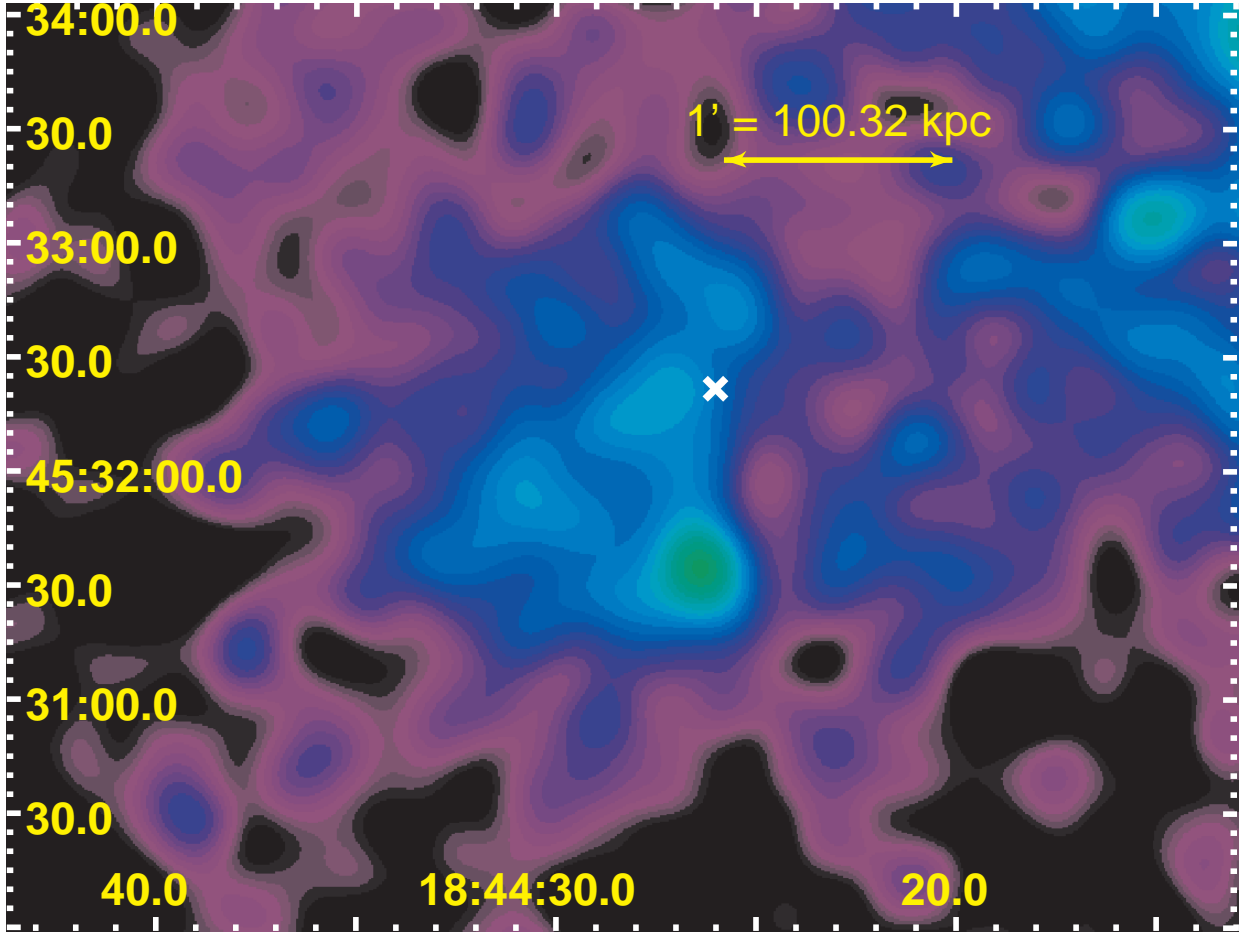}
\caption{Adaptively smoothed, background subtracted, exposure corrected
{\em Chandra}/ACIS-I image in the 0.5-2.0 keV band of the subcluster to the E of 3C 388.
The white X corresponds to the position of the optical/IR
galaxy 2MASX J18442603+4532219.}\label{subcluster}
\end{figure}

An X-ray enhancement 5$'$ ($\sim$500 kpc) East of the cluster center
may be a subcluster that is either falling toward 3C 388 or has already
passed through the cluster core.
Unfortunately the ACIS-I chip gaps lie directly across the bridge between
the primary cluster and the sub-cluster, complicating an
assessment of their relationship.
An adaptively smoothed, background subtracted, exposure corrected
image of this subcluster in the 0.5-2.0 keV band is shown in Figure~\ref{subcluster}.
The morphology of this region is roughly triangular with diffuse peaks
of X-ray emission at each vertex.  One vertex is pointed away from the cluster center.
The only known galaxy within 1$'$ of this subcluster is 2MASX J18442603+4532219, which
lies at the western boundary of the X-ray enhancement as shown in
Figure~\ref{subcluster}, and is not associated with any of the
X-ray peaks.  There is no redshift reported in the literature for this object,
although it appears to be two merging galaxies in the 2MASS images.
This pair is approximately 2 magnitudes fainter than the host galaxy
of 3C 388 in the K band.
The temperature of the gas in the subcluster is 1.9$^{+2.1}_{-0.7}$ keV assuming
$Z$=0.5$Z_\odot$, and the X-ray luminosity is 1.7$\times$10$^{42}$ 
ergs s$^{-1}$ (within a 50$''$ radius circle) in the 0.1-10.0 keV band assuming 
that it lies at the distance of the 3C 388 cluster.
The quality of the data is not sufficient to conclusively determine whether this
subcluster is falling toward 3C 388, is an unrelated foreground or background object,
or has already passed through the cluster core.  
If the latter hypothesis is correct, the subcluster passed through the core
$\sim$500 Myrs ago (assuming a velocity of 1000 km s$^{-1}$), roughly an
order of magnitude larger than the age of the radio source (see below).  In this
case, the merger will have deposited considerable energy in the core.

\section{Interpretation}

Our measurements of the thermodynamic parameters
of the ICM permit us to make quantitative statements about the dynamics and
energetics of the lobe inflation.
Based on considerations of the jet/counterjet brightness, \citet{lea01} estimate
the jet makes an angle of $\sim$50$^\circ$ with respect to the line
of sight.  This conclusion is supported by several lines of
evidence including the axial ratio of the lobes and approximate
equality of the Faraday depths of the lobes.
For consistency, we use this estimate here.
Our results below are not strongly dependent to this choice.

\subsection{Hydrodynamics of buoyant hot bubbles}

The surface brightness decrements and X-shaped morphology of the X-ray emission around
the radio lobes indicate that the lobes have evacuated cavities
in the gas.  It is believed that the lobes of all radio galaxies are greatly
overpressurized relative to the ambient medium early in their life \citep{rey01}.
However, as discussed above, there is no evidence for
sharp surface brightness discontinuities or large changes in
gas temperature that would indicate strong shocks.
If thermal conduction in the ICM is efficient (i.e. at or near the
Spitzer value), shocks may be nearly isothermal and difficult to detect \citep{fab05a,fab05b}.
However, {\em Chandra} observations the contact discontinuity
in cluster `cold-fronts' \citep{vik01} and
the high Mach number shock around the SW radio lobe of Centaurus A \citep{kra03,kra05b}
suggest that the thermal conduction (and viscosity) of the ICM is orders
of magnitude below the Spitzer value.
Therefore, we conclude that the lobes are inflating transonically or subsonically (i.e. $M\leq$1).
This also suggests that the lobe are at most only moderately overpressurized
relative to the ambient medium (factor of 2 at most \citep{lan89}).
Once the lobes reach pressure equilibrium with the ICM,
their motion will be driven by buoyancy \citep{chu02}.
We note that if the lobes are expanding supersonically, our conclusions
regarding the energetics and dynamics of the lobes would be strengthened
as we will have underestimated the energy of the outburst and overestimated
the age of the lobes.

We assume that the lobes are in the buoyant stage to estimate their velocity
of expansion and the age of the source.
The expansion of the lobes was faster in the past, so our estimate of the
age of the source will be an upper limit.
We approximate the age of the bubble as the radius of the
bubble divided by the buoyant velocity.
The velocity of a buoyant bubble can be estimated by equating the buoyancy force
with the drag (ram pressure).  In this case, the velocity of the
bubble is given by
\begin{equation}
v=\sqrt{g \frac{V}{S}\frac{2}{C}},
\end{equation}
where $g$ is the acceleration of gravity, $V$ and $S$ are the volume and
cross-section of the bubble, respectively, and $C$ is the drag coefficient.
The drag coefficient is a function of both the geometry and the
Reynolds number, $R$, and is of order unity for $R$ in the range of 10$^{3}$ to 10$^{5}$.
We assume that there is a large difference between the mass density of the lobe, $\rho_{lobe}$,
and the ambient medium, $\rho_{ICM}$ (i.e. $\rho_{lobe}$/$\rho_{ICM}<<$1).

\citet{chu02} and \citet{bru02} estimate the drag coefficient
for the buoyant radio bubbles in M87 using hydrodynamical arguments and
numerical simulations.  They conclude that $C\sim$0.7 and
$v\sim$0.6$\times c_s$.
A similar calculation using the parameters of 3C 388 shows that
$v\sim$0.9$\times c_s$.
This is probably an overestimate for three reasons.  First, the drag
coefficient increases considerably for flows near Mach 1 (e.g. at $M$=1, $C$ is
roughly 4 times larger than for $M<<$1) due to the dissipative
effect of weak shocks.  
Second, if the viscosity of the ISM is the Spitzer value \citep{spi62},
the Reynolds number of the flow in 3C 388 is $\sim$70, considerably lower than
the flows considered by \citet{chu02} and \citet{bru02}.
The drag coefficient increases as $R$ decreases for $R<$10$^{3}$ (Figure 34 of \citet{lan89}).
For a solid sphere rising buoyantly in a fluid with $R$=70, the drag
coefficient, $C$, is 1.2.
Third, the simple model of balancing the buoyant force and
the drag is probably not a good approximation.  The radius of the lobes
is of the order of the distance of the lobes from the nucleus, and
the effect of the compressibility of the lobes
and the ambient gas, must be considered.
The drag would be larger than estimated above in a detailed
model that included a more complex geometry and compressibility of the gas.
We therefore conservatively estimate $v\sim$0.5$c_s$, with an uncertainty of
perhaps a factor of two, and thus estimate the age of the lobes as $\leq$65 Myrs.

We model each lobe as a cylinder with the axis parallel to
and perpendicular to the jet axis for the East and West lobes,
respectively, to estimate the volume of the lobes.
Assuming the same geometry as above (i.e. the lobes lie at an
angle of 50$^\circ$ with respect to the line
of sight) and the best-fit
$\beta$-model distribution for the gas, we estimate the `bubble' enthalpy
(that is, the work done on the gas by the inflation of the lobe
plus the internal energy of the lobe, 4$pV$ for $\gamma$=4/3) to be 5.2 and 
6.9$\times$10$^{59}$ ergs for the E and W lobes, respectively.
The total mechanical energy of the AGN outburst therefore is 1.2$\times$10$^{60}$ ergs,
of which 3.0$\times$10$^{59}$ ergs has already gone into work
done on the cluster gas.

The inflation of the lobes has already added a significant amount
of energy to the cluster core.  The total thermal energy of the gas
within the radius of the radio lobes is 1.8$\times$10$^{60}$ ergs.
The work done on this gas by the lobes is 3.0$\times$10$^{59}$ ergs, or
$\sim$0.8 keV per particle.  
This estimate probably overestimates the heating per particle because
the heat input has already spread beyond the region occupied
by the lobes.
If the lobes are currently evolving buoyantly, and hence subsonically, as
we have argued above, the effect of the input heat will have propagated into
the ambient ISM at the sound speed of the gas, or roughly twice the inflation
velocity of the lobe.  The energy has been input to a much larger volume
of gas.  Assuming the input energy has been added uniformly to the gas within twice
the radius of the lobes, $2\times R_{lobe}$, the amount of energy added is only 0.3 keV per particle.

It is likely that we are witnessing the ongoing
quenching of a cluster cooling flow by a nuclear outburst.
The temperature of the gas in the cluster core must have
been $\sim$0.3 keV cooler before the outburst than at present (assuming the heat has been
uniformly distributed).
For the current 3.5 keV gas temperature at the cluster
center, the gas cooling time, $\tau_c\sim E_{t}/\epsilon$, where $E_{t}$ is the
thermal energy density of the gas and $\epsilon$ is the emissivity, is $\sim$ 1 Gyr.
If the gas in the cluster core were cooler, and therefore presumably denser,
before the radio outburst, its radiative cooling time would have been considerably
shorter, and it is likely that there was a significant accumulation of cooler gas.
It is also probable that
the cluster core is still reacting to the energy deposited by the
inflation of the lobes.  The gas in the cluster core is probably not
in hydrostatic equilibrium in the gravitational potential.
Depending on where the energy from the inflation of the
lobes is ultimately deposited (i.e. in the cooling core or farther out
in the halo), a temperature inversion
may be created (i.e. the core may become hotter than the halo) in which
case the gas would become convectively unstable.

The fossil group ESO 3060170 may be an example of a much later stage
of a similar evolutionary process \citep{sun04}.  {\em Chandra} and
XMM-Newton observations demonstrated that there is no group scale
cooling core in this group, although the radiative cooling time
of the gas was much less than the Hubble time.  It was suggested
that a powerful radio outburst with total energy of a few $\times$10$^{59}$
ergs (i.e. roughly comparable to what we have observed in 3C 388) may have
reheated the cooling gas and quenched the cooling flow.  There is
no evidence for cavities in the gas in ESO 3060170, so any bubbles
have presumably risen buoyantly into the halo and dispersed.

The amount of heating in the cluster could ultimately be significantly
larger than $\sim$0.3 keV per particle for three reasons.
First, the initial stage of the bubble inflation was probably
supersonic \citep{rey01}. A considerable amount of energy above
the bubble enthalpy could have been added to the gas during this stage.
Second, the bubble enthalpy will eventually be converted to
thermal energy of the gas as the bubble rises in the atmosphere \citep{chu02}.
The enthalpy of the bubble is converted to kinetic energy of the
gas as it falls in behind the rising bubble, then ultimately thermalizes 
into internal energy of the gas.  The scale height to which the bubble will rise
and within which the energy deposition takes place is uncertain and depends
on several factors including how much ambient gas is entrained in the buoyantly
rising bubble and whether or not the bubble remains adiabatic.
In the case of Hydra A and Perseus A, the bubbles are likely to deposit
at least half of their enthalpy into thermal energy of the gas within
the cooling radius \citep{nul02,bir04}.
Finally, the jets and lobes are currently still powered by the active nucleus.
Depending on how much longer the outburst lasts, the energy deposited
in the gas could be a factor of a few larger than the current bubble
enthalpy.

The rate at which the mechanical power is deposited into
the gas is the $pV$ work divided by the inflation time of the lobe or
$\sim$4$\times$10$^{44}$ ergs s$^{-1}$.
This is two orders of magnitude greater than the current radiative 
power output of the active nucleus, and
is not atypical compared to other
radio galaxies/cluster cavities that have been studied in detail.
A sample of 18 X-ray cavities in clusters, groups, and galaxies
has been studied from data in the {\em Chandra} archive \citep{bir04} 
where it was found that the ratio between the mechanical power of the 
outburst and the radiative power of the AGN or the radio luminosity
of the lobes varied between factors of ten to several hundred.
The large difference between the mechanical power and the radio power
of radio galaxies was predicted on theoretical grounds
\citep{dey93,bic97}.  As discussed by \citet{owen00} and \citet{bir04}, the radio
power of a radio galaxy is not a good indicator of the mechanical
power of the outburst.

The energy input by the inflation of the lobes is more than sufficient to
balance radiative losses in the core.
We define the cooling radius, $r_{rad}$ as the distance from the nucleus where the
cooling time is 5 Gyrs (33.4 kpc for 3C 388).  This is roughly the distance at which radiative losses
must be balanced by another source of energy.  The cooling time of 5 Gyrs
roughly corresponds to the time between major merger events.  
Within this radius, the X-ray luminosity is 
2.2$\times$ 10$^{43}$ ergs s$^{-1}$ (unabsorbed) in the 0.1-10.0 keV band.
This is roughly a factor of 20 less than the lower limit to the rate
of mechanical energy deposition of the current outburst.  If the duty cycle of outbursts of the
observed magnitude is only 5\%, the accumulation of cooling gas at the
cluster center will be quenched.
Our results are consistent with
hydrodynamic simulations of cyclical nuclear outbursts in galaxy 
clusters. \citet{dal04} demonstrate that repetitive outbursts with
mechanical power similar to 3C 388 with
a duty cycle of 5-10\% can suppress the formation of cooling gas.

If the interpretation of \citet{roe94} for 3C 388 is correct and the differences
in radio spectral index of the radio lobes indicates multiple outbursts,
the picture is somewhat more complicated.  Assuming that 3C 388 is not
an X-shaped radio galaxy seen in projection \citep{lea84,kra05},
the more recent radio outburst has `caught-up' with the older
outburst and implies near transonic velocities.  The older outburst
must be evolving buoyantly, and if the more recent outburst
has displaced the older bubble, then the lobes are inflating transonically and
perhaps supersonically.
This would have several important implications.  First,
our estimate of the evolutionary timescale of the bubble would be reduced 
by a factor of four or more, with a consequent increase in
the rate of mechanical energy deposition by the same factor. 
The discrepancy between the current luminosity of the AGN and the
time averaged rate of mechanical work to inflate the lobe becomes even
larger.  Second, we will have considerably underestimated the total energy deposited into
the cooling flow.
The older outburst must also have deposited a similar
amount of energy into the ICM (assuming a similar volume).
In addition, transonic or supersonic inflation velocities imply that
the hot ISM has been shock heated.

\subsection{Internal pressure of radio lobe}

The equipartition pressure of the diffuse regions of the radio lobes
is $\sim$4$\times$10$^{-12}$ dyn cm$^{-2}$ based on GHz frequency radio measurements.
Using the best-fit models of the $\beta$ profile, and assuming that
the jet/counterjet makes an angle of $\sim$50$^\circ$ with respect to the line
of sight, we estimate the thermal gas pressure at the approximate
position of the lobes to be $\sim$7.0$\times$10$^{-11}$ dyn cm$^{-2}$, more than an order of magnitude
larger than the minimum pressure of the lobes.
We conclusively confirm the results of \citet{hard00} and \citet{lea01} (based
on ROSAT HRI observations) that the radio lobes are greatly underpressurized relative to
the ambient medium.  
It is not feasible that the lobes are so underpressurized relative to the gas as they would
be crushed in a sound crossing time if this were the
case.  Either the equipartition assumption is incorrect,
or something else dominates the pressure of the lobes.

A significant difference between the equipartition pressure of lobes and
jets in FR I radio galaxies and the pressure of the ambient medium
has now been observed in a large number of sources and appears
to be a common phenomenon (e.g. \citet{cro03,dun04}).
However, X-ray observations of inverse-Compton scattering of CMB photon
from relativistic electrons in the radio lobes of FR II radio galaxies in
poorer environments than 3C 388
have demonstrated that the relativistic electrons responsible
for the GHz radio emission are not far out of equipartition
with the magnetic field \citep{hard02,cro05}.  
Based on morphology and radio power, 3C 388 is classified as
an FR II, yet the large difference between the equipartition pressure
and the ambient medium is typical of FR Is.
If the lobes of 3C 388 are not in equipartition but the GHz emitting
particles and lobe magnetic field are in pressure
equilibrium with the gas, either there is an excess of particles,
or the field is stronger than the equipartition value.  If the lobe
is particle dominated, the magnetic field must be approximately
10\% of the equipartition value to maintain
pressure balance with the gas, which is well outside the distribution
inferred from lobe IC emission by \citet{cro04}.
On the other hand, it is possible the lobes are magnetically dominated,
but this is typically difficult to arrange.

There are several possible sources for this `missing'
pressure including lower energy relativistic electrons ($\gamma\sim$10,
which would radiate at kHz frequencies and be unobservable)
or protons, partial filling factor of the relativistic gas, and
entrained thermal protons.
Each of these hypotheses has significant problems, however.
If a population of lower energy relativistic electrons dominates the pressure,
the magnetic field should come into equipartition with this.  
The fact that the field is close to
equipartition with GHz-emitting charged particle population means that
additional energetically dominant particle distributions are unlikely
\citep{cro05}.
The latter two possibilities are more difficult to rule out observationally.
It is possible that the radio emitting plasma is clumpy, but
as \citet{lea01} point out, the filling factor would have
to be extremely low ($\sim$1-2\%) to retain equipartition.
It is also possible that thermal gas has been entrained into the lobe
via Kelvin-Helmholtz instabilities.  The visibility of the X-ray
cavity suggests that such plasma, if present, must be considerably hotter and less dense
than the ambient ICM.  The entrained gas could then be heated via thermal
conduction by the relativistic plasma in the lobe.  No such hot gas has been
detected in any radio lobe, and the temperature of this gas would have to
be  high ($>$15 keV) for it not to be visible in nearby radio
galaxies with {\em Chandra} \citep{nul02,sch02}.
\citet{maz02} discovered a hot ($T$$\sim$7.5 keV) bubble of gas
in the cluster MKW 3s that appears as a surface brightness deficit in
the {\em Chandra} image.  There is no radio emission associated with
this X-ray bubble, and whether it is the result of the entrainment and
heating of the ICM gas during the inflation of a radio lobe
is unknown.

\section{Conclusions}

We detect cavities in the X-ray emission from the cluster
gas coincident with the radio lobes of 3C 388.
There is no evidence for temperature or surface brightness
discontinuities suggestive of strong shocks
in the gas.  Thus the lobes must be evolving transonically
or subsonically.  
The work done by the inflation of the lobes on the gas is at least
0.3 keV per particle out to twice the radius of the lobe.
The bubble enthalpy, when ultimately converted to thermal energy of the gas,
will increase this by a factor of a few.
The rate of mechanical energy deposition by the inflation of the
lobes is at least twenty times greater than the radiative power
of the ICM within the cooling radius.  Thus, repeated outbursts of
similar power with
a duty cycle of only a few percent are more than capable of suppressing
the accumulation of cool gas at the core of this cluster of galaxies.
Because of the large difference between the mechanical power input into
the gas by the inflation of the lobes and the radiative losses, this
is one of the best examples of the quenching of cooling gas by
a nuclear outburst.

The equipartition pressure
of the lobes is more than an order of magnitude less than the
pressure of the ambient medium.  The pressure of the lobes must
therefore be dominated by something other than the relativistic particles
emitting at GHz radio frequencies.
In this regard, the lobes of 3C 388 are more similar to the plumes and lobes
of FR I radio galaxies than to the lobes of FR II sources (at least
those in poor environments).

\acknowledgements

This work was supported by NASA contracts NAS8-38248, NAS8-39073,
the Royal Society, the Chandra X-ray Center, and the Smithsonian Institution.
We thank the anonymous referee for comments that improved this paper.

\clearpage

\clearpage

\clearpage

\clearpage

\clearpage

\clearpage

\clearpage

\clearpage

\clearpage

\clearpage

\clearpage

\begin{table}
\begin{center}
\begin{tabular}{|c|c|c|}\hline\hline
Region & Temperature (keV) & Abundance \\ \hline
  1    & 2.8$_{-0.7}^{+1.2}$ &  0.7$_{-0.6}^{+2.8}$  \\ \hline
  2    & 3.1$_{-0.8}^{+1.0}$ &  0.3$_{-0.3}^{+1.0}$  \\ \hline
  3    & 3.4$_{-0.7}^{+1.1}$ &  0.6$_{-0.4}^{+1.4}$  \\ \hline
\end{tabular}
\caption{Summary of best fit temperatures and abundances (fraction
of $Z_\odot$) in three regions around 3C 388.  Uncertainties are
at 90\$ confidence.  See text (section 3.1) and
Figure~\ref{central} for description of regions.}\label{spectab}
\end{center}
\end{table}

\end{document}